\documentclass[twocolumn,preprintnumbers,amsmath,amssymb,superscriptaddress,prl]{revtex4}

\usepackage{graphicx}
\usepackage{dcolumn}
\usepackage{bm}
\usepackage{tikz} 
\usepackage{footmisc}
\usepackage{float}
\newcommand*\circled[1]{\tikz[baseline=(char.base)]{
            \node[shape=circle,draw,inner sep=1pt] (char) {#1};}}
\begin{document}
\bibliographystyle{prsty}

\title{Cardiac Mechano-Electrical Dynamical Instability}
%
%
%

\author{L. D. Weise}
\affiliation{Department of Physics and Astronomy, Ghent University, Krijgslaan
281, S9, Ghent, 9000, Belgium}
\affiliation{Theoretical Biology, Utrecht University, Padualaan 8,
Utrecht, 3584 CH, The Netherlands}
\author{A. V. Panfilov}
\affiliation{Department of Physics and Astronomy, Ghent University, Krijgslaan
281, S9, Ghent, 9000, Belgium}
\date{August 14, 2019}

\begin{abstract}
In a computational study we reveal a novel dynamical instability of excitation
waves in the heart muscle.
The instability manifests itself as gradual local increase in the duration of
the action potential which causes formation and hypermeandering of spiral
waves.
The mechanism is caused by stretch-activated currents that cause wave
front-tail collisions and beat to beat elongation of the action potential
duration due to biexcitability.
%
%
%
%
%
%
We discuss the importance of the instability for the onset and dynamics of
cardiac arrhythmias.
%
%
\end{abstract}

\maketitle

%
%
Spiral waves of excitation have been found in many biological, physical, and
chemical systems~\cite{Alles:76, gerisch_stadienspezifische_1965,
agladze_waves_2000}.
Spiral waves emerge in excitable media after wave break, a temporary, local
block of wave propagation, after which the wave curls around it's back forming
a spiral~\cite{Belousov,ZaZh70}.
%
%
%
The emergence of spiral waves in the heart muscle causes life-threatening
cardiac arrhythmias~\cite{WinfStr:84}.
Therefore it is of great interest to understand mechanisms that cause wave
break in the heart.
%
%
Break formation can be a result of anatomical
heterogeneity~\cite{zipes_sudden_1998} or dynamical instability.
The most studied dynamical instability in cardiac tissue is ``alternans'' which
can occur via various mechanisms~\cite{garfinkel_eight_2007}.
Alternans manifests in as a beat-to-beat alternation in the duration of action
potentials (short-long-short), which grows in time and may result in wave break
formation.
%

%
The heart's contractions are governed by electrical waves of excitation.
Conversely, its deformation affects the excitation processes of the
cardiomyocytes, which is called ``mechano-electrical feedback'' (MEF).
MEF has been shown to be able to cause, but also to abolish dangerous cardiac
arrhythmias \cite{Kohl:99}.
%
%
However, so far no dynamical instability which is caused by MEF has been
identified yet.
%
%
%
%
%
%
%
%
%
%
%
%
%
In this letter we report the finding of such a mechano-electrical dynamical
instability (MEDI) in a model for human cardiac tissue.
%

%
%
Our method couples an ionic model for human epicardial
myocytes~\cite{TenTusscher:06}, with a discrete mechanical model for cardiac
tissue~\cite{Weise:11}, and a model for excitation-contraction
coupling~\cite{Niederer.bj06,Niederer.pbmb08} adjusted to human cardiac
tissue~\cite{keldermann_electromechanical_2010}.
The propagation of nonlinear waves of electrical excitation in cardiac tissue
is modeled via a reaction-diffusion equation for the transmembrane
potential~$V$
  \begin{equation}
  \label{Eq1}
  \dfrac{\partial V}{\partial t} = D \Delta V -
  \dfrac{I_{ion}+I_{sac}}{C_{m}},\\
  \end{equation}
with membrane capacitance density~$C_m=2.0\;\mu F/cm^{2}$ and
diffusivity~$D_{ij}=\delta_{ij}\times~1.54\;cm^{2}/s$.
At boundaries of the medium no-flux boundary conditions are used ($\nabla
V=0$).
%
The transmembrane ion current~$I_{ion}$ is modeled by various time- and
voltage-dependent ion channels~\cite{HH4:52}.
%
%
%
The finite difference mesh for the explicit Euler integration (space
step~$0.25\;mm$ and time step~$0.02\;ms$) of Eq.\eqref{Eq1} is coupled to a
square lattice of mass points connected with springs (see Figure~$1$
in~\cite{weise_discrete_2013}).
Excitation waves trigger a contraction of the
tissue~\cite{weise_discrete_2013}.
To solve the mechanical model we assumed elastostatics, and used Verlet
integration~\cite{Verlet:67}.
To model MEF we use a linear, time-independent model for stretch-activated
currents
  \begin{equation}
  \label{eqn_Is-bp}
  {I_{sac}} =
  G_{s}\dfrac{\left(\lambda-1\right)}{\left(\lambda_{max}-1\right)}\left(V-E_{s}\right),
  \mbox{ for } \lambda>1
  \end{equation}
%
%
where~$\lambda$ is a measure for local dilatation: strain in one-dimensional
(1D) simulations, and square root of the relative area change of a
quadrilateral formed by direct neighboring mass points (see Figure~$1$
in~\cite{weise_discrete_2013}) in two-dimensional (2D) simulations.
Parameter~$\lambda_{max}$ is maximal normalized sarcomere length which
we chose as~$\lambda_{max}=1.1$ as in~\cite{keldermann_electromechanical_2010}.
$G_{s}$ is the maximal conductance,~$E_{s}$ the reversal potential of the
stretch activated channels.
$E_{s}$ was measured in a range from~$-20\;mV$
to~$0\;mV$~\cite{kohl_stretch-induced_1999,Skouibine.mb00}.
We set~$E_{s}=0\;mV$.
%
%
We vary~$G_s$ in the reported range from~$0$
to~$100\;S/F$~\cite{Kohl:99,Kohl.cjc98}.
%
%
Following similar studies~\cite{PhysRevLett.108.228104,PanfKeldNash:05,
Panfilov.pnas07} we fixed the boundaries of the model to mimic isovolumic
phases of the cardiac cycle.
Our 2D model relates to a thin slice of cardiac tissue with fixed boundaries.
%
%
For~1D simulations we assumed a constantly stretched cable ($\lambda =
\lambda_{max}$) and vary~$G_s$ in Eq.\eqref{eqn_Is-bp}.
%
%
%
%

\begin{figure}[ht]
\centerline{
\includegraphics[width=1.0\columnwidth,clip]{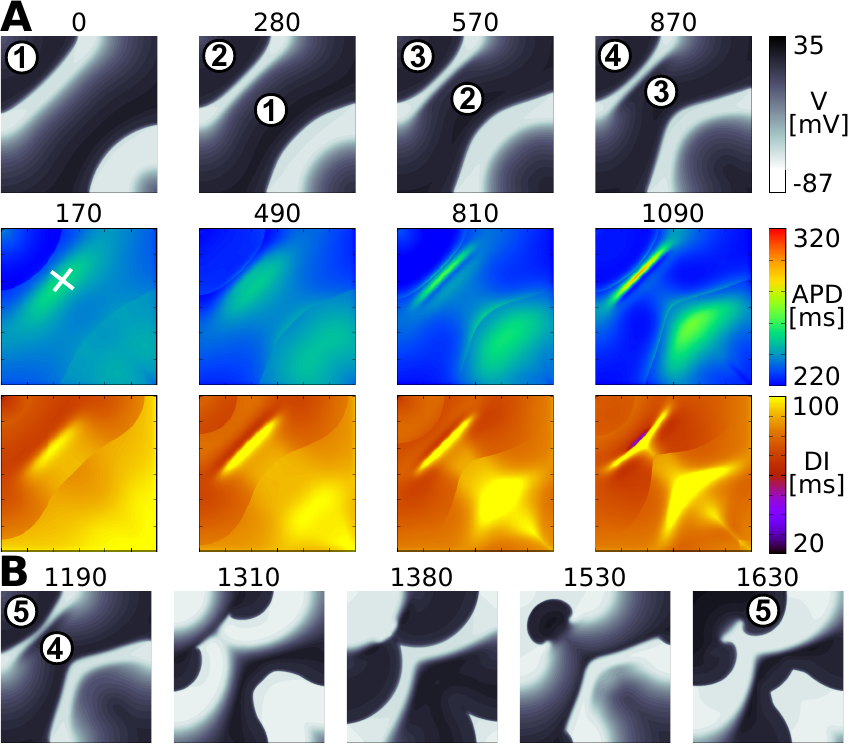}
}
\caption[]{MEDI.
Periodical wave initiation in left, upper corner:
ten waves with period~$340\;ms$, then seven waves with
period~$300\;ms$.
Five last waves (\circled{$\mathbf 1$}--\circled{$\mathbf 5$}) are shown.
(A) \circled{$\mathbf 1$}--\circled{$\mathbf 4$}: Wave front-back
collisions and emergence of spatial heterogeneity.
Upper panel: Transmembrane voltage (V); snapshots are taken, when wave
front-back collision happens.
White cross indicates region where DI is minimal and MEDI develops.
Middle panel: Action potential duration (APD).
Lower panel: Diastolic interval (DI).
(B) Wavebreak and spiral formation.
Time~[$ms$] after first wave front-back collision is shown above a snapshot.
Side length of model~$15\;cm$.
$G_s = 50\;S/F$.
}
\label{Figure1}
\end{figure}

%
%
Figure~\ref{Figure1} and supplemental movie~\footnote{LINK TO SUPPLEMENTAL
MOVIE} shows development of MEDI under periodic stimulation of cardiac tissue.
%
%
First, we stimulate the tissue with a constant period of~$340\;ms$ and observe
stable wave propagation.
However, when we decrease the stimulation period to~$300\;ms$, we see
development of MEDI.
In Figure~\ref{Figure1}A, top we show wave front-back collisions before wave
break happens (compare supplemental movies~\footnote{LINK TO SUPPLEMENTAL
MOVIE} and figure~\footnote{LINK TO SUPPLEMENTAL FIGURE}).
%
%
%
We see next, that at this location APD~\footnote{APD and DI are recorded
at~$-60\;mV$} gradually increases, while DI decreases (Figure~\ref{Figure1}A),
until wave break occurs evolving to two counter-rotating spiral waves
(Figure~\ref{Figure1}B).
Note also, that in contrast to alternans instability MEDI occurs for longer
stimulation periods than classical APD alternans~\cite{TenTusscher:06}, and
does not involve alternations between long and short action potentials.
%
%
%

\begin{figure}[ht]
\centerline{
\includegraphics[width=1.0\columnwidth,clip]{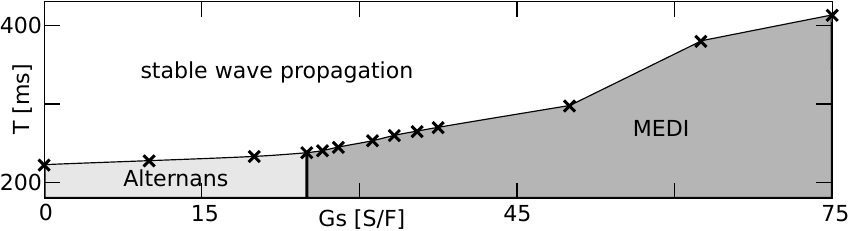}
}
\caption[]{Wave patterning as a function of period of
stimulation (T) and~$G_s$.
Robustness of MEDI. 
Above the line: stable wave propagation.
Thick line: onset of dynamical instability resulting in wave break.
Crosses: measurements. 
Dark grey area: MEDI.
Light grey area: alternans instability.
Waves were started with an initial period of~$0.5\;s$.
After ten stimulations the period was decreased by~$2.5\;ms$, and this was
repeated until wave break happened.
Setup as in Figure~\ref{Figure1} was used.
}
\label{Figure2}
\end{figure}

%
How robust is MEDI against the change of model parameters?
To answer this question we used the setup shown in Figure~\ref{Figure1};
however, slowly decreased the period of stimulation for different~$G_s$.
We show the results in Figure~\ref{Figure2}.
%
We see from Figure~\ref{Figure2} that MEDI occurs in a large parametric space
of~$G_s$ and period of stimulation~$T$.
%
%

%
We will now explain the mechanism of MEDI.
First, we need to explain what causes wave front-back collisions.
%
%
We reported in~\cite{PhysRevLett.119.108101} that external dynamic stretching
of cardiac tissue causes acceleration of a wave front and promotes wave
front-back collision.
Here we have a similar situation; however, the stretch is not caused by an
external mechanical load, but the contraction is caused by the excitation wave
itself.
The mechanism is the same: stretch causes~$I_{sac}$ which accelerates the wave
front causing wave front-back collision.
However, why does the APD at the collision position grow from wave to wave (see
Figure~\ref{Figure1}A)?
This is counter-intuitive, because classical restitution theory predicts that
collision (short DI) should produce shorter APD.
However, classical restitution theory cannot be applied here as the collision
is a non-stationary spatio-temporal process which cannot be reproduced by
a periodic stimulation of cardiac cell.
Here one needs to consider the interaction of the wavefront with the waveback
of the preceeding wave~\footnote{compare APD vs. DI plot for dynamic
restitution protocol and collisions LINK TO SI}.
%
%
%
%
%
%

\begin{figure}[ht]
\centerline{
\includegraphics[width=1.0\columnwidth,clip]{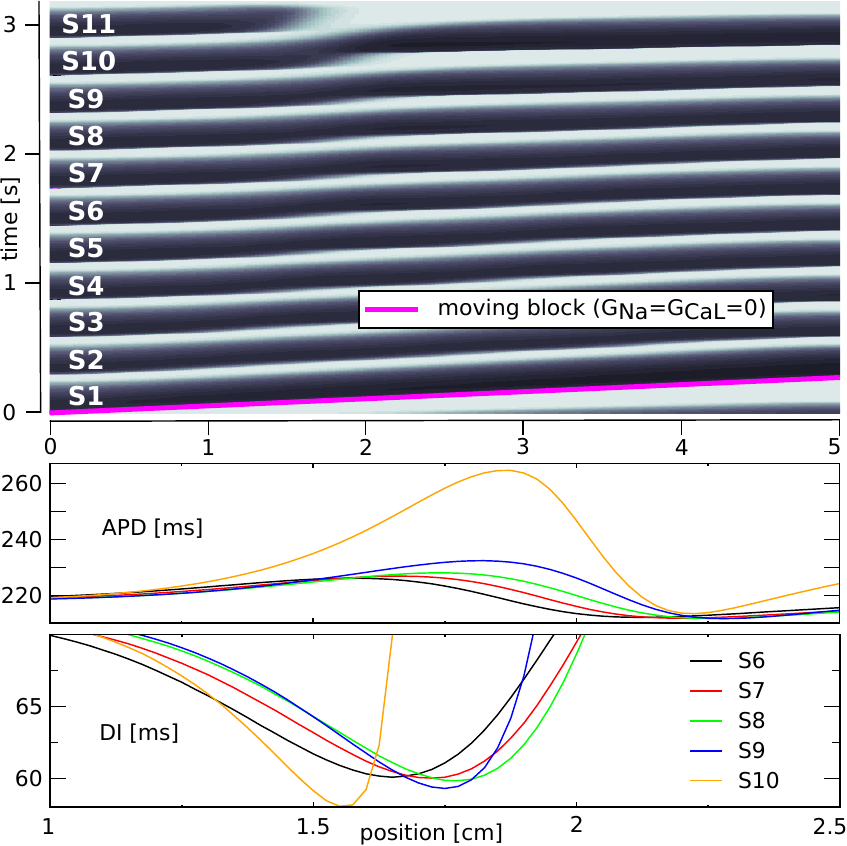}
}
\caption[]{Stepwise APD increase and wave block in constantly
stretched cable.
Top: Time-space plot.
S1 is forced to constant CV of~$17.65\;cm/s$ by obstacle (thick line).
Middle panel: APD of a wave train (S6--S10) {\it vs} position.
Lower panel: DI of a wave train (S6--S10) {\it vs} position.
%
Waves were started with period~$290\;ms$ in~$5\;cm$ long constantly stretched
cable ($\lambda = \lambda_{max}$).
Cable was prepared with twelve waves before S1 wave.
$G_s = 10.5\;S/F$.
}
\label{Figure3}
\end{figure}

%
To systematically study wave front-back collisions we developed a special
electrophysiological setup.
%
%
In this setup we use a moving obstacle in a cable to control the velocity of a
first wave (S1) (see also Figure~3\ in~\cite{PhysRevLett.119.108101}), and
initiate a wave train (see Figure~\ref{Figure3}).
To study the effect of MEF we assume the fiber to be constantly stretched
to~$\lambda_{max}$, thus~$I_{sac}=G_s(V-E_s)$.
By changing the velocity of the moving obstacle we can systematically vary the
degree of wave front-back interaction during collision.
%
%
We can see that we can reproduce the observed MEDI in this setup by choosing a
corresponding value of the forced velocity (Figure~\ref{Figure3}).
In particular, for forced velocity of~$17.65\;cm/s$ we see collisions of
successive waves, gradual increase in APD (Figure~\ref{Figure3}B) and gradual
decrease of the DI at the collision position (Figure~\ref{Figure3}C).
This process closely resembles the instability in the 2D system (compare
Figure~\ref{Figure3} and Figure~\ref{Figure1}).
%
%

\begin{figure}[ht]
\centerline{
\includegraphics[width=1.0\columnwidth,clip]{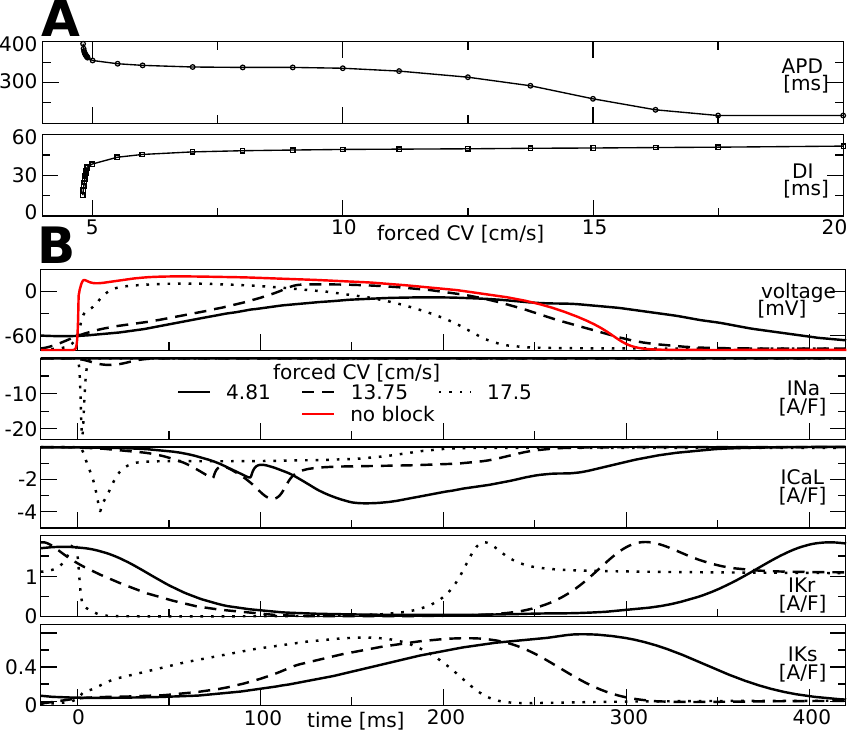}
}
\caption[]{APD elongation is caused by biexcitability.
Electrophysiological observables at position of S2-S1 collision (when DI
between S2 and S1 is minimal) {\it vs} forced CV.
(A) Top: APD {\it vs} forced CV.
Bottom: DI {\it vs} forced CV.
%
%
(B) Electrophysiological observables of S2 wave~{\it vs} time at S2-S1
collision position as a function of forced CV.
Time is shifted to start of upstroke, when~$V=-60\;mV$.
%
%
Top: transmembrane voltage V.
Other panels: strongest polarizing and depolarizing currents
Waves were started with period~$0.5\;s$ in~$20\;cm$ long cable.
%
Cable was prepared with twelve waves before S1 wave.
$G_s = 10.5\;S/F$.
}
\label{Figure4}
\end{figure}

%
%
We performed additional simulations to systematically study front-tail
interactions by letting two waves (S1 and S2) collide for different velocities
of the moving obstacle.
Figure~\ref{Figure4} shows wave characteristics at the collision point, i.e.
when SI between S1 and S2 is minimal.
Note, that this point has a different location for different forced CV.
We see from Figure~\ref{Figure4}A that, as expected, lower forced CV results in
closer front-tail interaction (DI decreases).
However, we also see that such decrease in DI results in unexpected increase in
APD which is counter to normal APD restitution, where shorter DI results in
shorter APD.
Such abnormal dependency, can, in our view, explain the observed MEDI.
Indeed, for periodic forcing with a period T, DI=T-APD, thus
increase in APD will result in decrease in DI.
However, if decrease in DI will produce longer APD, as in the case of
Fig.~\ref{Figure4}A this longer APD will produce shorter DI and will further
increase APD and thus result in its gradual growth, what we see in
Figure~\ref{Figure1} and Figure~\ref{Figure3}.
%
%
How does this ``abnormal´´ APD(DI) dependency emerge? 
We find that it is related to the phenomenon ``biexcitability''.
Under special conditions~\cite{vandersickel_study_2014,chang_bi-stable_2012}
bistable wave propagation can occur in the same tissue.
One type is fast propagation: characterized by a rapid sodium-driven upstroke
(INa) happening from the repolarized transmembrane potential, and the other is
a slow propagation where the upstroke is driven by L-type calcium current
(ICaL) from a depolarized potential when sodium channels are mostly inactivated
due to accommodation~\cite{qu_mechanisms_2015}.
%
%
%
%
%
We found earlier in the moving obstacle setup, that~$I_{sac}$ causes
biexcitability of the S2 wave which manifests in its oscillation between
sodium- (when S2 is distant from the S1 wave back) and calcium-driven upstroke
(during wave front-back collision)~\cite{PhysRevLett.119.108101}.
%
%
Here we find that such a transition of wave front propagation substantially
affects APD.
We can see (Figure~\ref{Figure4}B) that for a forced CV of~$17.5\;cm/s$,
where APD is~$219\;ms$ the upstroke of the action potential is steep, and driven
by sodium current (dotted lines).
%
%
However, for slower forced CV (straight and dashed lines) we can see that the
slope of the action potential becomes shallow, sodium current is absent,
and L-calcium current is the main depolarizing current.
The action potentials are also substantially longer,~$292\;ms$ for forced CV
of~$13.75\;cm/s$, and~$395\;ms$ for forced CV of~$4.81\;cm/s$.
%
%
%
We can explain the APD elongation by a combination of the longer transient of
the calcium current, and a delay of the repolarizing currents IKs and IKr for
slower forced CV happening as a consequence of the change of the propagation
type (Figure~\ref{Figure4}B lower panels).
%
%
We studied the importance of ICaL for the mechanism.
In the setup of Figure~\ref{Figure3} we found that
wavebreaks due to MEDI occur for forced velocities $[12.4;15.26]\;cm/s$.
However, if we block ICaL we did not observe MEDI; we could either see stable
wave propagation (forced CV~$>15.26\;cm/s$) or immediate block of S2 (forced
CV~$<15.26\;cm/s$), and no MEDI.
%
%
%
Thus we can conclude that ICaL and related to it biexcitability is a key part
of MEDI.
Overall we can explain the mechanism of MEDI as follows:
MEF due to stretch-activated currents increases the local velocity of the
wavefront which causes wave front-back collisions.
A wave front-back collision results in short DI, which
however causes longer APDs at the collision regions due to biexcitability.
Elongation of APD further decreases DI in the collision region which further
increases APD.
This positive feedback results in wave by wave increase of APD until the wave
front dissipates.
%
%

\begin{figure}[ht]
\centerline{
\includegraphics[width=1.0\columnwidth,clip]{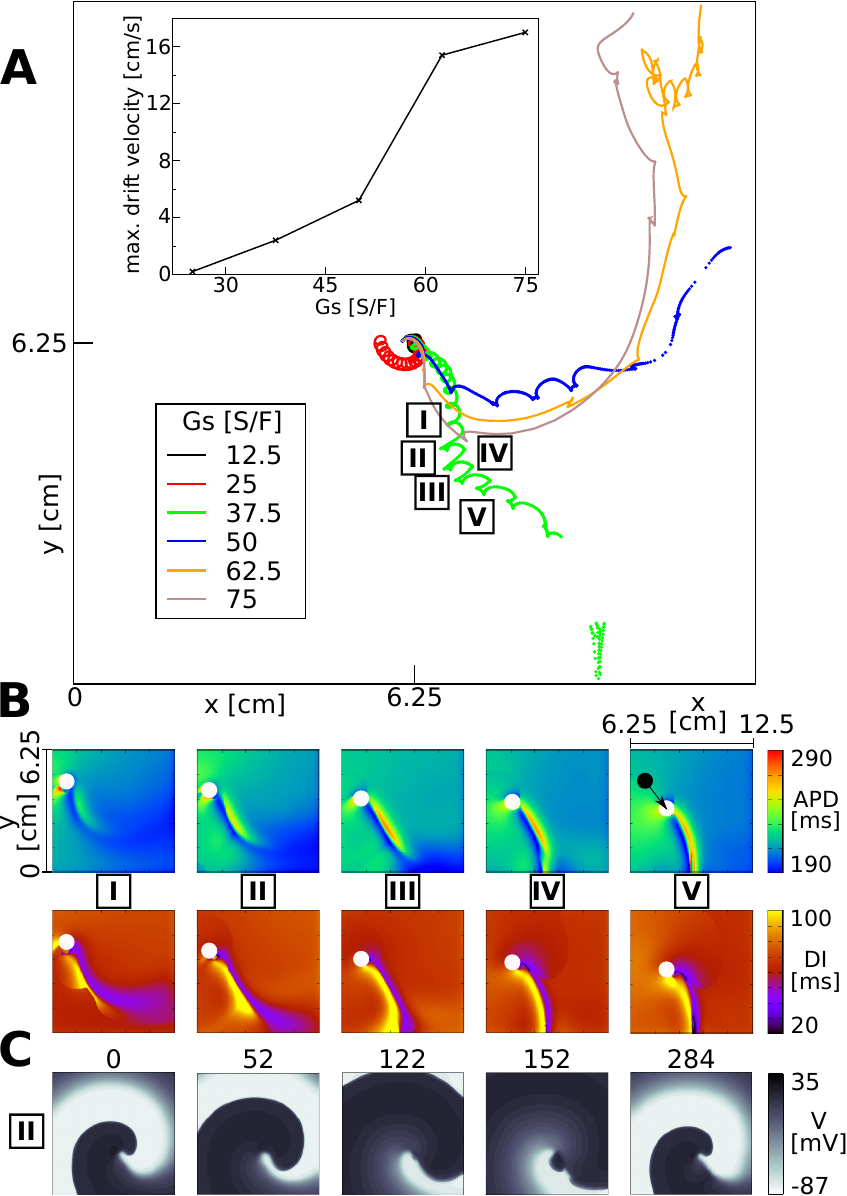}
}
\caption[]{MEDI causes rapid spiral drift.
(A) spiral tip trajectories as function of~$G_s$.
Inset: drift velocity as function of~$G_s$.
(B) APD and DI after wave back-front collisions show MEDI.
Lower, right quadrant of the medium is shown.
Location of spiral core is illustrated by a white dot.
Snapshots are taken for~$G_s=37.5\;S/F$ (compare subfigure A).
In snapshot~{\bf V} we illustrate position of spiral core of snapshot~{\bf I}
as black dot, and indicate drift direction with an arrow.
(C) Illustration of collision~{\bf II} (compare subfigure B) leading to
temporary wave block close to spiral tip.
Time [$ms$] (starting at~$2488\;ms$ simulation time) is shown above each
snapshot.
Length of medium~$12.5\;cm$.
Spiral was initiated in the medium without MEF~($G_s=0\;S/F$), let rotate
for~$2\;s$, then system was saved and used as starting point for simulations
with MEF.
}
\label{Figure5}
\end{figure}

%
How relevant is this mechanism for cardiac arrhythmias?
We showed in Figures~\ref{Figure1},~\ref{Figure2} that MEDI can lead to wave
break and creation of spiral waves in the heart.
Therefore MEDI may be relevant for the onset of cardiac arrhythmia.
%
%
%
Does MEDI also destabilize spiral waves?
We also studied what effect MEDI has on spiral wave dynamics.
We show the results in Figure~\ref{Figure5}.
%
We see that for~$G_s=12.5\;S/F$ the spiral wave has a circular core which is
the same as dynamics in absence of MEF.
However, for~$G_s=25\;S/F$ MEF causes meandering of the spiral wave on a
cycloidal trajectory.
This onset of meander can be explained by the resonant drift
theory~\cite{Gril:95} which predicts meandering under a periodical variation of
the excitability of the medium, which occurs here due to
MEF~\cite{Panfilov.pnas07, dierckx_theory_2015}.
However, for larger values of~$G_s$ we observe a hyper-meandering trajectory
(see the orange line).
Such hyper-meandering has substantial consequences for the type of arrhythmia,
as it induces polymorphic ventricular
tachycardia~\cite{gray_nonstationary_1995}.
We found that the maximal induced drift velocity is approximately one third of
the wave velocity (see inset in subfigure~\ref{Figure5}A, wave velocity is
ca.~$60\;cm/s$) which is too slow to cause ECG patterns similar to ventricular
fibrillation~\cite{Gray:95b}.
We found that the rapid spiral drift is caused by MEDI.
We illustrate it in Figure~\ref{Figure5} for the spiral core trajectory
$G_s=37.5\;S/F$ (green line), where we show collisions {\bf I}--{\bf V}.
In subfigure~\ref{Figure5}B we show that MEDI occurs.
%
%
Similar to Figure~\ref{Figure1}A we see that APD  gradually increases  (see
collisions~{\bf I}--{\bf III} in subfigure~\ref{Figure5}B).
We also see that it causes wave block close to the spiral wave tip which does
not result in full spiral breakup.
We observe that the spiral core drifts along the ``collision line'' (region where DI
is minimal and APD maximal).
We indicate the ``drift vector'' in subfigure~\ref{Figure5}B,{\bf V} as an arrow.

%
%
%
In this letter we reported on our finding of a novel dynamical instability
``MEDI'' for excitation waves in cardiac tissue.
MEDI emerges as a consequence of MEF and causes the formation and
hypermeandering of spiral waves.
These phenomena are relevant for the onset of cardiac arrhythmias.
MEDI causes wave break in a large range of conductivities of stretch-activated
channels and stimulation periods.
MEDI can occur for longer stimulation periods than the alternans instability.
It is difficult to formulate an analytical theory for MEDI.
This is, because it emerges from the interplay of the complex phenomena CV and
biexcitability that themself depend on the interplay of wave propagation and
MEF.
To explain the mechanism we studied the wave by wave increase of APD in
a simplified 1D setup (Figure~\ref{Figure4}), in which we disabled the
electromechanical coupling.
However, in the full model it is more complex, as APD affects also the
spatiotemporal strain distribution in the medium, and thus affects CV.
We see no easy way to analytically reduce the complexity of this
spatiotemporal problem.
We studied MEDI in a simplified 2D model for cardiac tissue.
As a next step it may be interesting to investigate the novel mechanism in more
detailed three-dimensional electromechanical models for the human
heart~\cite{trayanova_electromechanical_2011,
keldermann_electromechanical_2010}.
For example, it is important to test if the novel instability can cause a
breakup of vortices, as this is a key mechanism for sudden cardiac death.
It may be possible to experimentally study MEDI, for example by using
ultrasound-based strain imaging~\cite{christoph_electromechanical_2018} in
animal models.
It can also be interesting to design experiments similar to our forced CV setup
(compare Figure~\ref{Figure4}) using the optogenetics
approach~\cite{magome_photo-control_2011}.
%
%
%
%

%
%
{\it Acknowledgements:}
L.D.W. thanks the Deutsche Forschungsgemeinschaft for a research fellowship (Grant
No. WE 5519/1-1).
We thank Dr. Vadim Biktashev, Dr. Jan Kucera Dr. Hans Dierckx and Dr. Ivan
Kazbanov for valuable discussions.
We thank Dr. Paul Baron for critical comments on the manuscript.
%

\end{document}